\documentclass[11pt]{amsart}

\usepackage[margin=1in]{geometry}

\usepackage{amsmath,amsfonts,amsthm,mathrsfs}
\usepackage[unicode]{hyperref}
\usepackage{amssymb}
\usepackage{enumitem}
\usepackage{booktabs} 
\usepackage{caption} 
\usepackage[table]{xcolor} 

\usepackage{graphicx}
\usepackage{wrapfig}
\newcommand{\name}{\texttt{CausalEC}}
\newcommand{\namespace}{\texttt{CausalEC {}}}

\begin{document}

\title{C\MakeLowercase{omments on} ``C\MakeLowercase{ausal}EC: A C\MakeLowercase{ausally} C\MakeLowercase{onsistent} D\MakeLowercase{ata} S\MakeLowercase{torage} A\MakeLowercase{lgorithm} B\MakeLowercase{ased on} C\MakeLowercase{ross-}O\MakeLowercase{bject} E\MakeLowercase{rasure} C\MakeLowercase{oding}"
}
\author{R\MakeLowercase{amy} E. A\MakeLowercase{li}  
\\ E\MakeLowercase{mail}: \href{mailto:reali@usc.edu}{\MakeLowercase{ramy.ali@samsung.com} }} \thanks{Ramy E. Ali was with Penn State University. He is now with Samsung Electronics. This work was done prior to joining Samsung Electronics.}

\maketitle

\begin{abstract}
Cadambe and Lyu 2021 presents an erasure coding based algorithm called \name{} that ensures causal consistency based on cross-object erasure coding. This note shows that the algorithm presented in Cadambe and Lyu 2021 and the main ideas behind it are in essence the same as the algorithm developed in Lyu, Cadambe, Ali and Urgaonkar 2018. 

\end{abstract}

\section{Summary of ``Erasure coding based causally consistent distributed data storage'' \cite{2018causalec}}
First, I provide a description of the 2018's work \cite{2018causalec}. Specifically, I state the contributions, the guarantees ensured by the protocol, the techniques developed to ensure such guarantees and give pointers to the proofs.

\subsection{Contributions} 

This paper presents \name, which is a protocol that ensures a guarantee termed as \emph{causal consistency} (defined in page 3, 4th paragraph) based on a technique called \emph{erasure coding} (Definition 1 and Definition 2 in page 4) for the first time. Specifically, all prior works use a technique called replication or another technique called partial replication (paragraphs 1 and 2 in page 1). That is, this paper uses erasure coding to provide causal consistency for the first time. In particular, the paper developed the protocol based on a technique termed as ``inter-object coding'' (cross-object coding) for the first time as mentioned in the 3rd paragraph of page 2, and is explained in more detail below. The protocol also ensures more write, read and storage cost guarantees as discussed next.

\subsection{Guarantees} 

The inputs of the protocol are a collection of objects $\mathcal O$ and any arbitrary erasure code $C$, where $\mathcal O$ can have \textbf{any number of objects $K$} (the first paragraph of Section 5.1 in page 11). \emph{That is, this protocol is a general protocol that scales with the number of objects}. The protocol then ensures \textbf{causal consistency} (Theorem 4 in page 15) and that \textbf{all reads eventually return the same value} (Theorem 14 in page 20), and the following properties.
\begin{enumerate}
    \item \textbf{Write Guarantees}. The protocol ensures that every write operation is local (See for instance the first three paragraphs in page 9). 
    \item \textbf{Read Guarantees}. For the objects that are uncoded, the protocol ensures that the reads are local. For the other objects, \namespace contacts a set of remote servers to decode and  return the required value. This is based on the recovery sets concept (page 4, Definition 2). (See for instance the first three paragraphs in page 9 and the Proof of Lemma 9 in page 17, which states that server $s$ first tries to serve the read request locally. If there is no local data structure that can serve this read request at server $s$, then server $s$ contacts and waits for messages from other servers to serve this read request). 
\end{enumerate}
Moreover, the protocol ensures that every operation issued by a non-failing (non-halting) client terminates provided that the number of server failures is at most $f$ (Theorem 11 in page 19). That is, Theorem 11 shows that every write operation issued to a non-failing server $s$ terminates (Lemma 12 in page 19). In addition, every read operation issued by a non-failing client at a server $s$ terminates provided that at most $f$ servers are non-failing (Lemma 13 in page 19, this is equivalent to saying that as long as $s$ and the servers in at least one recovery set are non-failing). Finally, the storage cost at the stable state is equal to that guaranteed by the underlying erasure code (Theorem 16 in page 20).

\subsection{Techniques}

More specifically, we developed the protocol based on a technique that we termed as ``\textbf{inter-object coding}''. This is one of the main novel aspects of this work and the key ingredient that enabled developing this protocol. \emph{We have not written the definition of inter-object coding as a definition in the paper, but we have explained it in \textbf{Section 1 at page 2 in the 3rd paragraph}}. In particular, the prior works that use erasure coding to ensure other guarantees (they do not ensure causal consistency) use a technique known as ``intra-object coding'' in which every object is split into multiple parts and each server stores a function of the object. As explained at the end of page 2, with intra-object erasure coding, no server stores an object value in its entirety and, therefore, it is impossible for a read to be local. In the inter-object coding technique, however, the object is not split before using erasure coding. \textbf{Hence,  inter-object coding is essential in ensuring the Write and Read Guarantees explained above.}  In this technique, the servers also have to \textbf{store history} of the objects in lists unlike all prior works that ensure causal consistency (See the first paragraph of page 6 and the first paragraph in page 12 for instance). 

\subsection{Protocol} The protocol is provided in Section 4.2 (page 6 to page 11) for a simplified setting (with $4$ servers).  The protocol consists of the following sub-protocols
\begin{enumerate}
    \item Client protocol (page 6, at the beginning of Section 4.2), which is straightforward, and
    \item Server protocol (Fig. 1 through Fig. 5), which consists of two sub-protocols as follows.
    \begin{enumerate}
        \item Data node protocol (Fig. 2, Fig. 3, end of page 7, page 8 and page 9).
        \item Parity node protocol (Fig. 4, Fig. 5 and page 9 to page 11).
    \end{enumerate}
\end{enumerate}

The general protocol (Fig. 6 to Fig. 9), for any number of servers and any number of objects, is described in Section 5 (page 10 to page 13). The general protocol also consists of the following sub-protocols
\begin{enumerate}
    \item Client protocol (page 12, 1st paragraph in Section 5.2), which is straightforward, and
    \item Server protocol (Fig. 6 to Fig. 9 from page 11 to page 15). This protocol is organized as follows. 
    \begin{enumerate}
        \item Fig. 6 (page 12) describes the initial states and the actions of the server.
        \item Fig. 7 (page 13) describes a part of the server protocol that is concerned with the input actions from clients.
        \item Fig. 8 (page 14) describes a part of the server protocol that is concerned  with the input actions from the other servers.
        \item Fig. 9 (page 16) finally describes the last part of the server protocol, which is concerned with the internal actions. 
    \end{enumerate}
\end{enumerate}
\textbf{The input of the protocol is a collection of objects $\mathcal O$ of any size} as stated in Section 5.1 (the first paragraph of Section 5.1 page 11). \\ These objects are encoded using an erasure code $C$ as explained in Section 5.1 (See the first paragraph of Section 5.1 at page 11, the third paragraph of page 12, the last paragraph of page 13 and Fig. 8). Since the code is applied across many objects as explained in page 11 in the first paragraph of Section 5.1, this means that \textbf{inter-object coding} is used. A key ingredient also of the protocol is that some servers \textbf{store history of the objects in a list} (first paragraph of page 12, Fig. 6, first paragraph of page 13, Fig. 7, Fig. 8 and Fig. 9). This does not happen in the prior works that ensure causal consistency. This history is cleared after some time through a Garbage Collection action (the first paragraph of page 12 and also at the end of Fig. 9). 

\subsection{Proofs} 

\textbf{It is first important to note that the proofs do not assume a specific number of objects. That is, the claims are proved in general for any number of objects and servers.} The proof of correctness is provided in Section 6.1. Specifically, this section is organized as follows. Section 6.1 (Theorem 4) proves that the causal consistency guarantee is ensured by the protocol. The proof starts by proving some auxiliary lemmas first (Lemma 6, 8, 9 and 10) and then using them to prove Theorem 4 as follows.
    \begin{enumerate}
        \item Lemma 6  (page 15), which shows that every write has a unique tag (See Definition 5 in page 15). This lemma is similar to Lemma 5.3 in page 19 of \cite{2021causalec}.
        \item Lemma 8  (pages 15, 16, 17), This lemma shows that if there is a path from an operation 1 to an  operation 2 in the graph of execution (See Definition 7 in page 15), then $\mathrm{timestamp \ of \ operation \ 1} \leq \mathrm{timestamp \ of \ operation \ 2}$. 
        \item Lemma 9  (page 17). This lemma shows that the value returned by a read is the value with the highest tag. This lemma is similar to Lemma 5.5 in page 20 of \cite{2021causalec}.
        \item Lemma 10 (page 18). This lemma shows that the tags of each encoded object in the pending reading list (\textrm{ReadL}) are the same.
   \end{enumerate}
The liveness proofs are provided in Section 6.2. In particular, the following results are proved. 
\begin{enumerate}
    \item Theorem 11 (page 19) shows that the write and read operations issued to a non-failing server always terminate. The proof of Theorem 11 is based upon Lemma 12 (page 19) and Lemma 13 (page 19) as follows.  
    \begin{enumerate}
        \item Lemma 12 shows that the \textbf{write operations} issued to a non-failing server always terminate.
        \item Lemma 13 shows that the \textbf{read operations} issued to a non-failing server always terminate provided that at most $f$ servers are non-failing (at least one recovery set of servers are non-failing).
    \end{enumerate}
    \item Theorem 14 (page 20) shows that all read operations eventually return the same value. The proof is based on Lemma 15 (page 20) which shows that the tags of two distinct values of the same object are always comparable. 
\end{enumerate}
Finally, Theorem 16 shows that the lists of the servers will be eventually empty at some point. That is, the servers may not need to store history after this point. This point is known as the ``stable state''. At this point, the storage cost of every server is equal to that guaranteed by the underlying erasure code. The proof of this theorem is not complete, but it follows from Theorem 14 as mentioned in page 20.  

\section{Summary of ``CausalEC: A Causally Consistent Data Storage Algorithm based on Cross-Object Erasure Coding'' \cite{2021causalec}}
Next, I provide a description of the 2021's work. Similar to the description of the 2018's work, I state the contributions, the guarantees ensured by the protocol, the techniques developed to ensure such guarantees and give pointers to the proofs. 

\subsection{Contributions}
This paper also presents \name, which ensures \emph{causal consistency} ( Definition 1 in page 8) based on the \emph{erasure coding} technique ( Definition 2 in page 9). As mentioned in Section 1 (3rd paragraph page 2), the use of erasure coding has not been explored before in the context of causal consistency. In particular, the paper developed the protocol based on a technique termed as  ``cross-object coding''. As explained in Section 1 (last paragraph of page 2), the object is not partitioned (split) in this technique. That is, this technique is the same  as the inter-object coding technique considered in \cite{2018causalec} to develop a causally consistent protocol. 
\subsection{Guarantees}

The inputs of the protocol are a collection of objects $K$ objects and any arbitrary erasure code $C$ (first paragraph in Section 4 in page 10). The protocol then ensures \textbf{causal consistency} (Theorem 5.1 in page 19) and that \textbf{all reads eventually return the same value} (Theorem 5.4 in page 23) and the following properties.
\begin{enumerate}
    \item \textbf{Write Guarantees}. The protocol ensures that every write operation is local (1st paragraph in Section 4.2.1 in page 12). 
    \item \textbf{Read Guarantees}. For the objects that are uncoded, the protocol ensures that the reads are local. For the other objects, \namespace contacts a set of remote servers to decode and  return the required value. This is based on the recovery sets concept (Definition 3 in Section 3.3 (page 9), see also the first two paragraphs in Section 4.2.1 in page 12).  
\end{enumerate}
Moreover, the protocol ensures that every write operation issued by a non-failing client terminates (Theorem 5.2 in page 20) and that every read operation at server $s$ always terminates provided that $s$ and one recovery set of servers are non-failing (Theorem 5.3 in page 22). Finally, the storage cost at the stable state is equal to that guaranteed by the underlying erasure code (Theorem 5.5 in page 23).

\subsection{Techniques}
The protocol is based on a technique termed as ``\textbf{cross-object coding}''. This is the same technique as the inter-object coding considered in \cite{2018causalec}, but it is just termed as cross-object coding here instead.   \emph{Cross-object coding is not written as a definition in the paper, but it is explained in \textbf{Section 1 at page 2 in the last paragraph}}. As explained, the object is not partitioned (split) before using erasure coding in this technique. \textbf{Hence,  cross-object coding is essential in ensuring the Write and Read Guarantees explained above.}   As proposed in \cite{2018causalec}, the servers also have to \textbf{store history} of the objects in lists (see page 11 for instance).

\subsection{Protocol}
The protocol is provided in Section 4 (page 10 to page 18). The protocol consists of the following sub-protocols
\begin{enumerate}
    \item Client protocol (at the beginning of Section 4.2 in page 12), which is straightforward, and 
    \item Server protocol (Fig. 3, Algorithms 1, 2 and 3). This protocol is organized as follows. 
    \begin{enumerate}
        \item Fig. 3 (page 12) describes the initial states and the actions of the server. This is very similar to Fig. 6 (page 12) in \cite{2018causalec}.
        \item Algorithm 1 (Fig. 4 in page 13) describes a part of the server protocol concerned with the input actions from clients. This is similar to Fig. 7 in \cite{2018causalec}.
        \item Algorithm 2 (page 14) describes a part of the server protocol that is concerned  with the input actions from the other servers. This is similar to Fig. 8 (page 14) in \cite{2018causalec}. 
        \item Algorithm 3 (page 15) describes the last part of the server protocol, which is concerned with the internal actions. This is similar to Fig. 9 (page 16) in \cite{2018causalec}.
    \end{enumerate}
\end{enumerate}
The input of the protocol is a collection of $K$ objects as described in Section 4.2 (2nd paragraph of Section 4.2 at page 12).\\
These objects are encoded using an erasure code. Storing codeword symbols means that an erasure code is used to generate such codeword symbols. Specifically, in the 4th paragraph of page 11, it is mentioned that each server stores codeword symbols (see also Algorithm 3 starting from line 14 that describes the ``Encoding'' and page 17). Since the code is applied across many objects (See for instance Section 3.3 in page 9), this means that \textbf{inter-object coding} is used. 

As in the protocol of \cite{2018causalec}, a key ingredient also of this protocol is that the servers \textbf{store history of the objects in a list} (See for instance the last paragraph in page 11, first two paragraphs of Section 4.2.1, ...). This history is cleared after some time through a Garbage Collection action (the last paragraph of page 17, the first paragraph of page 18 and the end of Algorithm 3).

\subsection{Proofs}
The proofs are provided in Section 5. Section 5.2 (Theorem 5.1 in page 19) shows that the causal consistency guarantee is ensured by the protocol. The proof of Theorem 5.1 depends on the following Lemmas.
\begin{enumerate}
    \item Lemma 5.4 (page 19).
    \item Lemma 5.5 (page 20). This lemma is similar to Lemma 9 in \cite{2018causalec}.
\end{enumerate}
The liveness proofs are then provided in Section 5.3. Specifically, the following results are shown.
\begin{enumerate}
    \item Theorem 5.2 (page 20) shows that the \textbf{write operations} issued to a non-halting (non-failing) server always eventually terminate. This is the same as Lemma  12 in \cite{2018causalec}. 
    \item Theorem 5.3 (page 22) shows that every \textbf{read operation} issued by a non-halting client at a server $s$ terminates as long as $s$ and the servers in one recovery set are non-halting. The proof of this theorem is based upon Lemmas 5.6, 5.7, 5.8 and 5.9 (all are in page 21). This is the same as Lemma  13 in \cite{2018causalec}. 
    \item Theorem 5.4 (page 23) shows that all read operations eventually return the same value. This is the same as Theorem  14 in \cite{2018causalec}.  
\end{enumerate}
Finally, Theorem 5.5 (page 23) shows that the storage cost is equal to that guaranteed by the underlying erasure code. This is the same as  Theorem 16 in \cite{2018causalec}. The proof depends on Lemma 5.12 (page 23), which shows that the list will eventually be empty at some point known as the stable state.  

\noindent Finally, I summarize the sub-protocols and the results of the two works \cite{2018causalec} and \cite{2021causalec} in Table \ref{table:protocol} and Table \ref{table:comparison}.

\begin{table}[htb!]
  \rowcolors{2}{gray!25}{white}
   \centering
  \begin{tabular}{ccc}
    \rowcolor{gray!50}
    Sub-protocol & 2018's work \cite{2018causalec}  & 2021's work \cite{2021causalec}  \\
     Server States &  Fig. 6  (page 12) & Fig. 3 (page 12) \\
    Input Actions from Clients & Fig. 7 (page 13) & Algorithm 1 (page 13) \\
    Input Actions from Other Servers & Fig. 8 (page 14) & Algorithm 2 (page 14) \\
    Internal Actions & Fig. 9 (page 16) & Algorithm 3 (page 15)
  \end{tabular}
  \caption{Summary of the  protocols of  \cite{2018causalec} and \cite{2021causalec}.}
  \label{table:protocol}
\end{table}

\begin{table}[htb!]
  \rowcolors{2}{gray!25}{white}
   \centering
  \begin{tabular}{ccc}
    \rowcolor{gray!50}
    Guarantee & 2018's work \cite{2018causalec}  & 2021's work \cite{2021causalec}  \\
    Causal Consistency &  Theorem 4  & Theorem 5.1 \\
    Termination of Writes & Theorem 11 (Based on Lemma 12) & Theorem 5.2 \\
    Termination of Reads & Theorem 11 (Based on Lemma 13) & Theorem 5.3 \\
    Reads eventually return the same value & Theorem 14 & Theorem 5.4 \\
     Storage cost is same as underlying erasure code & Theorem 16 & Theorem 5.5
  \end{tabular}
  \caption{Summary of the  guarantees of  \cite{2018causalec} and \cite{2021causalec}.}
  \label{table:comparison}
\end{table}

\bibliographystyle{IEEEtran}
\bibliography{ref}
\end{document}